\documentstyle[aps,prl,epsf,graphicx,multicol]{revtex}

\begin{document}

\def\tende#1{\,\vtop{\ialign{##\crcr\rightarrowfill\crcr
\noalign{\kern-1pt\nointerlineskip}
\hskip3.pt${\scriptstyle #1}$\hskip3.pt\crcr}}\,}

\def\insertplot#1#2#3#4{\begin{minipage}{#2}
\vbox {\hbox to #1 {\vbox to #2 {\vfil%
\includegraphics{#4.ps}#3}}}
\end{minipage}}

\def\ins#1#2#3{\vbox to0pt{\kern-#2 \hbox{\kern#1 #3}\vss}\nointerlineskip}

\title{Phase fluctuations in superconductors:\\
from  Galilean invariant to  quantum $XY$ models}
\author{ L. Benfatto, A. Toschi, S. Caprara,  and  C. Castellani}

\address{Dipartimento di Fisica, Universit\`a di Roma ``La Sapienza''  and 
	Istituto Nazionale per la Fisica della Materia, \\
        Unit\`a di Roma 1, Piazzale Aldo Moro, 2, 00185 Roma, Italy}
\maketitle
\begin{abstract}
We analyze the corrections to the superfluid density due to phase 
fluctuations within both a continuum and a lattice model for $s$- and
$d$-wave superconductors. We expand the phase-only action beyond the Gaussian 
level and compare our results with the quantum $XY$ model both in the quantum 
and in the classical regime. We find new dynamic anharmonic vertices, absent 
in the quantum $XY$ model, which are responsible for the vanishing of the 
correction to the superfluid density at zero temperature in a continuum 
(Galilean invariant) model. Moreover the phase-fluctuation effects are 
reduced with respect to the XY model by a factor at least of order 
$1/(k_F\xi_0)^2$. 
\end{abstract}
\vspace{-0.1 cm}
\begin{multicols}{2} \global\columnwidth20.5pc

The issue of explaining the linear temperature dependence of the superfluid 
density $\rho_s$ in cuprate superconductors \cite{exp} renewed the interest 
on the effects of the phase fluctuations (PF) of the order parameter in the 
superconducting phase at low temperature. Indeed, both quasiparticles 
\cite{qp} and PF \cite{pf} in the classical limit have been suggested as 
alternative explanations for the observed depletion of $\rho_s$. Recently 
Refs. \cite{randeria,noi} considered the quantum effects on PF and showed 
that the classical limit is reached at a temperature $T_{cl}$ which is too 
high to account for the low-temperature linearity of $\rho_s$ in cuprates. 

A crucial  point in the analyses of PF effects is the choice of the effective 
model for the PF. In Refs. \cite{randeria,noi} the Gaussian action for the 
phase $\theta$ was derived microscopically, while the anharmonic 
(non-Gaussian) terms were obtained by expanding the $\cos(\theta_i-\theta_j)$ 
coupling term of a lattice quantum $XY$ model, derived by coarse-graining
the Gaussian action on the scale of the coherence length $\xi_0$, in powers 
of $(\theta_i-\theta_j)\sim \xi_0 |\nabla\theta|$. In Ref. \cite{kwon}, 
instead, also the anharmonic terms were derived microscopically, within a 
$d$-wave {\em continuum} BCS model. In this approach the interaction vertices 
for $\nabla \theta$  are determined by the the fermion loops and are smaller 
than the corresponding vertices of the $XY$ model, leading to one-loop 
corrections to $\rho_s$ which are much smaller than those of the quantum $XY$ 
model, both at $T=0$ and at $T>0$. However, they found a finite correction to 
$\rho_s$ even at $T=0$, which is expected within a lattice $XY$ model, but 
not within a continuum (Galilean invariant) model where $\rho_s$ equals the 
particle density $\rho$ at $T=0$ \cite{popov}. This indicates that something 
is missing in their analysis. Moreover their description of PF effects within 
a continuum model is too restrictive, as it neglects lattice effects, which 
are certainly relevant in cuprates. 

In this paper we give a detailed analysis of the one-loop correction to
$\rho_s$ due to PF within both a continuum and a lattice model for $s$- or 
$d$-wave pairing. We specifically consider the weak- to intermediate-coupling 
regime, even though at the end we also comment on the strong-coupling limit. 
We find that the microscopic derivation of the phase-only action, besides the 
{\it classical} (static) anharmonic terms 
$(\nabla \theta)^4$ considered in Ref. 
\cite{kwon}, introduces new third- and fourth-order {\it quantum} (dynamic) 
interaction terms which contain the time derivative of $\theta$. These 
quantum terms are absent in the quantum $XY$ model, where the dynamics only 
appears at the Gaussian level, and induce a correction to $\rho_s$ which 
cancels exactly, in the continuum case, the contribution due to the classical 
interaction, restoring  the equality $\rho_s=\rho$ at $T=0$. On the other 
hand, the same cancellation does not hold in the lattice case, in which 
$\rho_s$ equals the average kinetic energy at $T=0$. The inclusion of both 
classical and quantum interaction terms leads to a finite one-loop correction 
to $\rho_s$, which is however of order $ 1/(k_F\xi_0)^2$ with respect to the 
result of the quantum $XY$ model, $k_F$ being the Fermi wave vector. In the 
classical regime we find that the PF correction to $\rho_s$ is smaller than 
within the classical $XY$ model by the factor $\sim 1/(k_F\xi_0)^2$, for both 
the continuum and lattice model. The reduction of the PF effects for 
$ k_F\xi_0 \gg 1$ is made even more pronounced in the continuum case by the 
inclusion of long-range Coulomb forces. The fact that the $XY$ model 
generically {\em overestimates} the PF effects further supports the claim of 
Refs. \cite{randeria,noi}, in which, even adopting the $XY$ model, it has 
been shown that the contribution of PF does not account for the linear 
temperature decrease of $\rho_s$ in cuprates and gives temperature-dependent 
small corrections when compared to the experimental data. 

We start with a continuum BCS action at a temperature $T=\beta^{-1}$
in $d$ dimensions, ${\cal S}=\int_0^\beta d \tau \cal L$, with
\begin{equation}
{\cal L}=\int d^dx \sum_\sigma c^+_\sigma \left (\partial_\tau-
\frac{\nabla^2}{2m}-\mu\right )c_\sigma+ H_I,
\label{mod}
\end{equation}
$H_I=-\frac{U}{\Omega} 
\sum_{{\bf k},{\bf k}',{\bf q}}\gamma_{\bf k}\gamma_{{\bf k}'}
c^+_{{\bf k}+\frac{{\bf q}}{2} \uparrow}
c^+_{-{\bf k}+\frac{{\bf q}}{2}  \downarrow} 
c_{-{\bf k}'+\frac{{\bf q}}{2} \downarrow} 
c_{{\bf k}'+\frac{{\bf q}}{2}  \uparrow}$. Here
$\Omega$ is the volume, $U>0$ is the pairing interaction strength and the 
factor $\gamma_{\bf k}$ controls the symmetry of the order parameter. 
In the following, unless explicitly indicated, we set $\hbar=k_B=1$. We 
perform the standard Hubbard-Stratonovich decoupling of $H_I$ and make the 
dependence on the phase $\theta$ of the complex order parameter 
$\Delta=|\Delta|{\rm e}^{i\theta}$ explicit by means of the gauge 
transformation  $c_\sigma \rightarrow c_\sigma {\rm e}^{i\theta/2}$ 
\cite{depalo,randeria,kwon,sharapov}. Then, after integrating out the 
fermions around the superconducting saddle-point solution, and neglecting the 
fluctuations of $|\Delta|$ \cite{notagap}, we obtain the effective 
action for PF,
${\cal S}_{eff}[\theta]={\rm Tr}\sum_{n=1}^\infty \frac{1}{n}(\Sigma G_0)^n$. 
$G_0$ is the mean-field Nambu Green function, and the self-energy matrix is
\begin{equation}
\Sigma=\left[ \frac{\dot \theta}{2}+\frac{(\nabla\theta)^2}{8m}\right]
\tau_3+\frac{i}{2m} 
(\nabla\theta \cdot \stackrel{\leftrightarrow}{\nabla})\tau_0 ,
\label{sigma}
\end{equation}
where $\tau_i$ are the Pauli matrices, and the operator 
$\stackrel{\leftrightarrow}{\nabla}
\equiv (\stackrel{\leftarrow}{\nabla}-\stackrel{\rightarrow}{\nabla})/2$ 
acts on $G_0$. In the following  we distinguish the ``bosonic'' 
contributions, generated by the $\tau_3$ term in (\ref{sigma}), analogous to 
those of a boson model in the presence of condensate, the ``fermionic'' 
contributions, generated by the $\tau_0$ term, and the ``mixed'' ones, 
obtained by combinations of $\tau_0$ and $\tau_3$ terms. Both bosonic and 
fermionic terms contribute to the Gaussian phase action which, in the 
hydrodynamic limit, reduces to the well-known form
\begin{equation}
{\cal S}_{eff}^G[\theta]=\frac{1}{8}\sum_{q} 
\left(\chi \omega_n^2+D {\bf q}^2\right)
\theta_q \theta_{-q},
\label{gauss}
\end{equation}
where $q=({\bf q}, i\omega_n)$, $\chi$ is the compressibility and 
$D(T)\equiv \hbar^2 \rho_s/m$ is the superfluid stiffness (in dimensional 
units). At Gaussian level the temperature dependence of $D$ is entirely due 
to the quasiparticle excitations, giving
$D(T)=\frac{\rho}{m}+\frac{1}{m^2\Omega}\sum_{\bf k}{\bf k}^2 f'(E_{\bf k})$. 
Here $f'(x)=-\beta {\rm e}^{\beta x}/({\rm e}^{\beta x}+1)^2$, 
$E_{\bf k}=\sqrt{\xi_{\bf k}^2+\Delta_{\bf k}^2}$  is the quasiparticle 
dispersion in the superconducting state, $\xi_{\bf k}={\bf k}^2/2m-\mu$ 
is the free-electron dispersion, and $\Delta_{\bf k}=|\Delta|\gamma_{\bf k}$ 
is the superconducting gap. 

\begin{figure}[h,t]
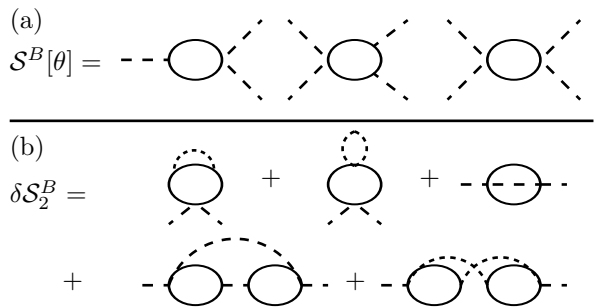

\begin{center}
\insertplot{220pt}{126pt}{
\ins{0pt}{125pt}{(a)}
\ins{0pt}{77pt}{(b)}
\ins{127pt}{25pt}{$+$}
\ins{95pt}{64pt}{+}
\ins{155pt}{64pt}{+}
\ins{0pt}{110pt}{${\cal S}^B[\theta]=$}
\ins{0pt}{60pt}{$\delta {\cal S}_2^B=$}
\ins{10pt}{25pt}{$\quad +$}
}
{figfase}
\end{center}
\caption{\small (a): Bosonic diagrams in the non-Gaussian phase-only action 
up to fourth order. The dashed line indicates a $\theta$ insertion,
the solid line represents $G_0$. The vertex with a single incoming line 
corresponds to the insertion of the $( {\dot \theta}/2) \tau_3$ term of Eq. 
(\ref{sigma}), the vertex with two incoming lines to the insertion of 
$((\nabla \theta)^2/8m) \tau_3$. (b): Bosonic one-loop corrections to the PF
Gaussian action.}
\end{figure}

In the present context, to make comparison with  Ref. \cite{kwon}, we
do not consider the effect of a dissipative term in Eq. (\ref{gauss})
for a $d$-wave superconductor, even though this term would be essential for 
the low-temperature behavior and for the estimate of the temperature $T_{cl}$  
\cite{noi}. We first consider a neutral system, while the long-range Coulomb 
forces will be considered later.

The one-loop PF corrections to $D(T=0)$ are induced by the anharmonic terms 
in $\theta$. As we discussed above, within the quantum $XY$ model the 
dynamics is introduced {\em only} at mean-field level adopting the PF 
Gaussian propagator $P(q)=4(\chi\omega_n^2+D{\bf q}^2)^{-1}$ (see Eq. 
(\ref{gauss})) \cite{randeria,noi}, and the one-loop 
correction to $D$, coming from the purely static interaction term 
$\propto \xi_0^2 D\sum_{\alpha=1}^d (\partial_\alpha \theta)^4$, is
\begin{equation}
\delta D_{XY}=-\frac{D\xi_0^2}{2}\frac{T}{d\Omega}\sum_{q}{\bf q}^2 P(q)
\tende{T\rightarrow 0}
-\frac{g}{\sqrt{\chi D}}\xi_0^2D,
\label{xy}
\end{equation}
where $g=\zeta^{d+1}/d(d+1)$, and $\zeta\simeq 1/\xi_0$ is the PF momentum 
cutoff \cite{notaxi}.

By contrast, we derive here the anharmonic terms by expanding 
${\cal S}_{eff}$ up to fourth order in $\theta$ and evaluate the one-loop 
corrections to the Gaussian action. All the fermionic and mixed terms 
generate one-loop corrections to $D$ which vanish or cancel each other at 
$T=0$, whatever is the symmetry of the gap, in agreement with the result 
previously discussed in Ref. \cite{kwon}, where however the contribution of 
the bosonic diagrams was incorrectly evaluated. Indeed, the bosonic 
$\dot \theta$ term in Eq. (\ref{sigma}) introduces a dynamic contribution in 
${\cal S}_{eff}$ also beyond the Gaussian level, as it is represented by the 
third- and fourth-order bosonic vertices depicted in Fig. 1a. In particular, 
the presence of a third-order quantum interaction vertex, neglected in Ref. 
\cite{kwon}, leads to separate cancellations between {\em both} fermionic or 
mixed {\em and} bosonic contributions. Let us consider the one-loop diagrams 
for the bosonic corrections $\delta {\cal S}_2^B$ in Fig. 1b: the first two 
are the self-energy corrections to $G_0$, which shift the particle density 
$\rho$ in the presence of PF, at fixed $\mu$. Alternatively, at fixed 
density, we take these contributions into account by shifting $\mu$ with 
respect to its mean-field value, in order to keep $\rho$ fixed. As a 
consequence, the only corrections to $D$ with respect to $\rho/m$ due to PF 
come from the last three diagrams in Fig. 1b,  i.e.
\begin{equation}
\delta D=-\frac{1}{8m^2d}\frac{T}{\Omega}\sum_{q}{\bf q}^2 P(q)
 \chi(q)\left[2-\omega_n^2 P(q)\chi(q) \right],
\label{dd}
\end{equation}
where  $\chi(q)$ is the density-density bubble, which gives the 
compressibility $\chi$ in the limit $\omega_n=0, {\bf q} \rightarrow 0$. 
Notice that in writing Eq. (\ref{dd}) we are relying on the fact that both 
the third- and fourth-order vertices needed to calculate $\delta D$ are 
expressed in terms of $\chi(q)$. This is the relevant consequence of the 
Galilean-invariant form of the bosonic $\tau_3$ term appearing in the 
self-energy (\ref{sigma}) \cite{depalo,thouless}. Evaluating Eq. (\ref{dd}) 
in the hydrodynamic limit, we get
\begin{equation}
\delta D=\frac{1}{dm^2\Omega}\sum_{\bf q} {\bf q}^2 b'(\varepsilon_{\bf q}) 
\tende{T\rightarrow 0} \, 0,
\label{corr}
\end{equation}
where $\varepsilon_{\bf q}=\sqrt{D/\chi}|{\bf q}|$ is the sound mode, and 
$b'(x)=-\beta{\rm e}^{\beta x}/({\rm e}^{\beta x}-1)^2$. The origin of this 
result is made clear by considering the analytic continuation of Eq. 
(\ref{dd}) to real frequencies, and summing the two terms. The pole at 
$\omega=-\varepsilon_{\bf q}<0$ (responsible for a finite contribution at 
$T=0$) is cancelled in favor of a double pole at 
$\omega=\varepsilon_{\bf q}>0$, leading to Eq. (\ref{corr}), where the 
standard Bogoljubov reduction of $\rho_s$ in a superfluid 
bosonic system is recognized. Thus, by fully including the dynamic structure 
of the interaction for the phase, we obtain that in a Galilean-invariant 
system $\rho_s=\rho$ at $T=0$ \cite{popov}.  

A further consequence of the microscopic derivation of the effective action
is a reduction of the strength ($\sim\chi$) of the static interaction term 
$(\nabla \theta)^4$ with respect to the $XY$ model, 
as already observed in Ref. \cite{kwon}. In the classical regime for PF only 
the term $\omega_n=0$ of (\ref{xy}) and (\ref{dd}) should be considered. 
Therefore in Eq. (\ref{dd}) only the correction to $D$ coming from the static 
interaction term $\propto (\chi/m^2) (\nabla \theta)^4$ survives, leading to 
$\delta D\approx -T\chi/d^2m^2 D  \xi_0^{d}$, {\em qualitatively} similar to 
the result of the classical $XY$ model, 
$\delta D_{XY}\approx -2T/d^2\xi_0^{d-2}$. Since, however, 
$\chi/m^2D \simeq 1/k_F^2$, we find that
\begin{equation}
\delta D/\delta D_{XY} \simeq \left(k_F\xi_0\right)^{-2},
\label{ratio}
\end{equation}
i.e., in the classical limit $\delta D$ is smaller than  within the $XY$ 
model, as far as $k_F^{-1}<\xi_0$. 

Let us now consider the effect of the Coulomb interaction between the 
electrons. At Gaussian level the density-density bubble $\chi(q)$ is dressed 
by the random-phase series of the Coulomb potential 
$V({\bf q})=\lambda e^2/|{\bf q}|^{d-1}$ (here $\lambda$ is a constant 
depending on the dimension $d$ and on the dielectric constant
$\epsilon_{\infty}$), and the sound mode $\varepsilon_{\bf q}$ of Eq. 
(\ref{corr}) is converted into the plasma mode $\omega_{\bf q}$ of the 
$d$-dimensional system. In deriving the anharmonic terms in ${\cal S}_{eff}$, 
we must now include the RPA density fluctuations in all the vertices. The 
one-loop corrections to $D$ are formally identical to Eqs. 
(\ref{dd})-(\ref{corr}), with $\chi\to\chi_{LR}$. Thus at $T=0$ we recover 
again the cancellations (\ref{corr}) of the bosonic diagrams, with 
$\varepsilon_{\bf q}\to\omega_{\bf q}$. At the same time, since 
$\chi_{LR}({\bf q},0)\approx 1/V({\bf q})$ vanishes as ${\bf q} \to 0$, the 
classical ($\omega_n =0$) term in Eq. (\ref{dd}) gives 
$\delta D_{LR} \approx -T/d(2d-1)Dm^2\lambda e^2 \xi_0^{2d-1}$, whereas 
the quantum $XY$ model leads to the same result as the neutral case. Thus we 
estimate
\begin{equation}
\delta D_{LR}/\delta D_{XY} \simeq (E_F/E_C)
\left(k_F\xi_0\right)^{-(d+1)},
\label{ratiolr}
\end{equation}
where $E_F$ is the Fermi energy and $E_C=\lambda k_F e^2$ is a characteristic 
Coulomb energy. While within the $XY$ model the Coulomb interaction modifies 
only the low-temperature behavior of $\rho_s$, within the continuum model it 
affects also the high-temperature classical regime. 

To extend the previous results to a lattice model, we rewrite Eq. 
(\ref{mod}) introducing a hopping $t$ between nearest-neighboring sites on a 
cubic lattice of spacing $a$, so that the free-electron dispersion is 
$\xi_{\bf k}=-2t\sum_{\alpha=1}^d \cos ak_\alpha-\mu$, and we obtain the 
generalization of Eq. (\ref{sigma}) to the lattice case. Since we find that 
the cancellation of fermionic and mixed one-loop corrections to $D$ at $T=0$ 
still holds, we focus on bosonic corrections only. Differently from the 
continuum case, each insertion of a spatial derivative of $\theta$ in the 
fermionic loops is associated to a same-order ${\bf k}$-derivative of  
$\xi_{\bf k}$. Therefore, the $(\nabla \theta)^2$ term in Eq. (\ref{sigma}) 
carries a factor 
${1\over 8}\Lambda_\alpha={1\over 8}{\partial^2 \xi_{\bf k}}/
{\partial k_\alpha^2}$. At Gaussian level, then,
$D(T=0)=\frac{1}{d\Omega}\sum_{{\bf k}, \alpha} \Lambda_\alpha 
(1-{\xi_{\bf k}}/{E_{\bf k}})$, 
which, in the nearest-neighbor cubic model, equals the average kinetic energy, 
rather than $\rho/m$. In the anharmonic action of Fig 1a the factor 
$\Lambda_\alpha$ appears in each vertex with two incoming $\theta$-lines: as 
a consequence, the first two diagrams of Fig. 1b, which give corrections to 
$\rho$ in the continuum case, are now corrections to $D$, which are different 
from (and therefore do not cancel with) those coming from the shift of $\mu$, 
at fixed $\rho$. Moreover, on the lattice, the coupling of the fermions to 
$\theta$ generates higher than second-order derivatives of $\theta$ in
(\ref{sigma}): in addition to the diagrams depicted in Fig. 1b we
must include also a diagram $\cal T$, with four derivatives of $\theta$ 
incoming in the same vertex, and a factor 
$\partial^4 \xi_{\bf k}/\partial k_\alpha^4$, proportional to the 
average kinetic energy. The resulting bosonic one-loop correction to $D$ is
\begin{eqnarray}
\delta &D&=-\frac{1}{8d} \frac{T}{\Omega}\sum_{q} P(q)\left \{ {\bf q}^2 
\left[ 2\chi_{EE}(q)+\chi_{EE} 
\right. \right. \nonumber\\ &+& \left.
(d-1){\tilde \chi}_{EE}+{\cal T}-d \chi^{-1}\chi_{\rho E}^2\right]
\nonumber\\
&- &
\left. \omega_n^2\left[\chi_{\rho E}^2{\bf q}^2P(q)- 2d \eta_{\rho E}(q)+
d\chi^{-1}\chi_{\rho E}\eta(q) \right]\right \}.
\label{ddl}
\end{eqnarray}
The $\chi_{ab}$ bubbles correspond to the insertion of one ($\chi_{\rho E}$) 
or two equal/different ($\chi_{E E}$/${\tilde \chi}_{EE}$) factors 
$\Lambda_\alpha$; $\eta_{\rho E}$ and $\eta$ are the bubbles with three $G_0$ 
lines and one or no factor $\Lambda_\alpha$ respectively. $\eta_{\rho E}$ 
corresponds to the first diagram of Fig. 1b. The first terms in the two 
square brackets are the lattice analogous of the two terms of the continuum 
case, Eq. (\ref{dd}), whereas the last contributions come from the shift of 
$\mu$ to keep $\rho$ fixed.

We perform the $\omega_n$ sum in Eq. (\ref{ddl}) with the PF Gaussian 
propagator $P(q)$, and retain the leading order in $\zeta\sim 1/\xi_0$,
calculating the $\chi_{ab}$ bubbles at zero incoming momentum, and carefully 
evaluating the small ${\bf q}$ limit for $\eta_{\rho E}$ and $\eta$. We thus 
obtain an overall correction to $D$ which is {\em finite} at $T=0$ in the 
lattice case,
\begin{eqnarray}
\delta D=&-&\frac{g}{4\sqrt{\chi D}}\left[ 3\chi_{EE}+(d-1)
{\tilde \chi}_{EE}+{\cal T}-2\chi^{-1}\chi_{\rho E}^2
\right. \nonumber \\&-&\left. 
d\chi^{-1}\chi_{\rho E}^2 \right]
-\frac{g(d+1)}{\zeta\chi}
\left[ \tilde \eta_{\rho E}-\chi^{-1}\chi_{\rho E}\tilde\eta
\right].
\label{ret}
\end{eqnarray}
At $T=0$, $\chi=\frac{1}{\Omega}\sum_{\bf k} N_{\bf k}$, with 
$N_{\bf k}=\Delta_{\bf k}^2/E_{\bf k}^3$; 
$\chi_{ab}=\frac{1}{\Omega}\sum_{{\bf k}}\Gamma_{ab}N_{\bf k}$, 
with $\Gamma_{\rho E}=\Lambda_\alpha$, $\Gamma_{E E}=\Lambda_\alpha^2$, 
$\tilde \Gamma_{E E}=\Lambda_\alpha\Lambda_\beta$  ($\alpha\neq \beta$); 
$\tilde\eta_{\rho E}=\frac{1}{\Omega}\sum_{{\bf k}}\Lambda_\alpha M_{\bf k}$ 
with $M_{\bf k}=\Delta_{\bf k}^2\xi_{\bf k}/E_{\bf k}^4$; 
$\tilde\eta=\frac{1}{\Omega}\sum_{{\bf k}}M_{\bf k}$;
${\cal T}\equiv a^2D(T=0)$. In the limit 
$a\rightarrow 0, \,t\rightarrow \infty$, keeping $2a^2t= 1/m$ finite, 
$\Lambda_\alpha \rightarrow 1/m$,   $\chi_{EE},{\tilde \chi}_{EE}
\to \chi/m^2$, $\chi_{\rho E}\to \chi/m$, 
$\tilde\eta_{\rho E}\to \tilde\eta/m$, while ${\cal T}=a^2D\rightarrow 0$. 
Thus, Eq. (\ref{ret}) recovers the continuum (Galilean-invariant) result  
$\delta D=0$. 

We next turn again to the issue of the comparison between $\delta D$ and 
$\delta D_{XY}$ in the lattice case. At $T=0$ Eq. (\ref{ret}) leads to an 
estimate of $\delta D/\delta D_{XY}$ of the same order of Eq. 
(\ref{ratio}), within a numerical factor. This is also true in the classical 
limit, both for the neutral and the charged system. As we discussed above, at 
high temperature only the static correction to $D$ contributes, with a 
coefficient controlled in the neutral case by $\chi_{EE}, \tilde\chi_{EE}$ 
and $\cal T$, which plays the same role (and has the same estimate) of 
$\chi/m^2$ in the continuum case. The presence of Coulomb forces does not 
change qualitatively this conclusion, and introduces only minor quantitative 
corrections, since the RPA expressions for  $\chi_{EE},\tilde\chi_{EE}$ have 
a {\em finite} limit for ${\bf q}\to 0$, contrary to the continuum case, e.g.,
$\chi_{EE}^{LR}=\chi_{EE}-\chi_{\rho E}^2V({\bf q})/[1+V({\bf q})\chi] \simeq 
(\chi_{EE}-\chi_{\rho E}^2/\chi)
+\chi_{\rho E}^2 |{\bf q}|^{d-1}/\chi\lambda e^2$. 
As a consequence, Eq. (\ref{ratio}) gives a proper estimate in the lattice 
case, even in the presence of long-range interactions. 

According to Eq. (\ref{ratio}) the PF-induced depletion of $\rho_s$ can be 
quite small at weak and intermediate coupling (particularly in the BCS limit) 
both in the quantum and in the classical regime, while it is of the order 
predicted by the $XY$ model in the strong-coupling limit $k_F\xi_0 \approx 1$.
In this case the corrections to $\rho_s$ due to PF are sizeable (even though 
less important than the quasiparticle contribution in determining the 
low-temperature dependence \cite{noi}). The evaluation of
$\chi_{ab},\tilde\eta_{\rho E}$ for $t/U\ll 1$ would lead to the conclusion 
that all contributions coming from the dynamic vertices are subleading with 
respect to those which arise from the static interaction 
$\sum_{\alpha=1}^d (\partial_\alpha \theta)^4$ \cite{dopo}. As a consequence, 
at strong coupling only the static interaction survives, analogously to what 
is assumed in the quantum $XY$ model. Moreover, in the case of $s$-wave 
pairing, the value of the coefficient of 
$\sum_{\alpha=1}^d (\partial_\alpha \theta)^4$ is exactly the same of the 
$XY$ model (\ref{xy}), with $\xi_0$ substituted by the lattice spacing $a$, 
in agreement with the strong-coupling expectation. For $d$-wave pairing the 
coefficient of the static interaction changes only by a numerical factor, in
units of $a$, with respect to the $XY$ model \cite{ondad}. 
 
However the situation is more involved, specifically for the $s$-wave pairing 
symmetry. Let us consider the negative-$U$ Hubbard model. In order to analyze 
the strong-coupling limit we need to include (i) the RPA fluctuations, also 
induced by $U$, in the particle-hole channel, and  (ii) the fluctuations of 
$|\Delta|\simeq(U/2)\sqrt{\rho(2-\rho)}$, which fluctuates because $\rho$ 
fluctuates \cite{depalo}. When this analysis is carried out the contributions 
from the dynamical vertices are {\em not} subleading, and those from the 
static vertices do not reproduce the $XY$ result by themselves. Nevertheless, 
at $T=0$ the inclusion of both dynamic and static corrections to $D$ leads 
again to $\delta D=\delta D_{XY}$, with $\xi_0$ substituted by $a$. This 
result holds also in the classical regime, even though only approximately 
and provided the particle density is not too small \cite{poi}.
\vskip 0.5truecm
\par\noindent
{\bf Acknowledgments}: We thank S. De Palo, C. Di Castro, M. Grilli, and A. 
Paramekanti, for many useful discussions and suggestions.

\end{multicols}


\begin{thebibliography}{99}
\bibitem{notagap} 
At low $T$, $|\Delta|$ couples to $\theta$ via its coupling to $\rho$. This 
coupling, which is irrelevant in the BCS limit, can be considered by 
including the fluctuations of $|\Delta|$ in the density bubbles.
\bibitem{notaxi} 
Spatial variations of $\theta(x)$ at distances smaller than $\xi_0$ must not 
be included. $\xi_0$ is well defined both in $s$- and $d$-wave 
superconductors through the spatial decay of the correlation function for 
$|\Delta|$.
\bibitem{ondad} 
At strong coupling in the $s$-wave case $\chi_{EE}$ and $D$ give the same 
contribution, while $\tilde \chi_{EE}$ is subleading. In the $d$-wave case 
$\tilde \chi_{EE}$ is not subleading, even if numerically less important 
than $\chi_{EE}$.
\bibitem{poi} 
The discussion of this issue (and of the relation between $\xi_0$ and $a$
at small density) is beyond the scope of the present paper, and 
will be reported elsewhere.
\end{thebibliography}
\end{document}